\documentclass[12pt]{article}

\begin{document}

\title{\textbf{Dissipative Viscous Cylindrical Collapse in $f(R)$ Gravity With Full Causal Approach}}

\author{G. Abbas \thanks{ghulamabbas@iub.edu.pk}
and H. Nazar \thanks{hammadnazar350@gmail.com}
\\Department of Mathematics The Islamia University\\ of Bahawalpur,
Bahawalpur, Pakistan.}
\date{}

\maketitle
\begin{abstract}
The plan of this study is to inspect the effects
of dynamics of dissipative gravitational collapse in cylindrical symmetric non-static spacetime
by using Misner-Sharp concept in context of metric $f(R)$ theory of gravity. For more generic isotropic fluid distribution of cylindrical
object by dissipative nature of dark source of the fluid due to energy matter tensor,the Misner-Sharp approach technique has been used to illustrate the heat flux with free radiating out flow, bulk and shear viscosity. Furthermore, dynamical equation have been associated with full
casual heat transportation equations in framework of $M\ddot{u}ller$-Israel-Stewart formalism. The present
study explain the effects of thermodynamics viscid/heat radiating coupling factors on gravitational collapse in
$M\ddot{u}ller$-Israel-Stewart notion and matches with the consequences of prior astrophysicists by excluding
such coefficients and viscosity variables.
\end{abstract}


\section{Introduction}

A provoking still inquisitive problem for cosmological, gravitating physics and along with
astronomical in general theory of relativity (GR) is to identify the concluding providence
of a incessant dissipative gravitating implosion. The firm arrangement of a gigantic object
endures longer the inmost pull of gravity is nullified by the external stress of the fissile
fuel at the centre of the object. Later, once the object has wakened its fissile fuel there
is no extended any thermodynamical scorching and there will be interminable gravitating implosion.
The configuration of the objects are made of explosive mass which incessantly gravitational and
is fascinated to its core due to gravitating interface of its particles. Later astrophysicist \cite{2}-\cite{4}
galactic perceptions like the investigation of type Ia supernova demonstrate that the extension
of the cosmos is rushing and its alleged this might be expected to a continuous positive cosmological
invariant. These perceptions have raised in thoughtfulness regarding the investigation of astral bodies
with cosmological invariant. Oppenheimer and Snyder \cite{5} are pioneers of this field of gravitating
source long decades ago. They made an activity stride in the field of gravitating collapse and examined
a nitty gritty work regarding the issue of gravitational collapse. At this order
following a quarter century, a brief sensible examination was accomplished by Misner and Sharp \cite{6}
with isotropic matter in the inside of a imploding object and as well as outside the object in vacuum.
Vaidya \cite {7} explain the gravitational collapse of radiating source.

Various paradigms have been opened to explicate the enigmatic behavior of dark energy (DE). Using the modified concepts of
gravity for instance $f(R)$, $f(R,T)$ and $f(G)$ etc. can be well elucidated for the expanding problem of universe.
Capozziello \cite {8} studied the metric $f(R)$ theory of gravity is the usual form of general relativity (GR) got
by presenting an uniformed function addiction on R (Ricci invariant). Sharif and Yousaf \cite{9} explored
the constancy of an imperfect fluid expansion region with non-static parameter of spherical symmetric model
in $f(R)$ framework. Mak and Harko \cite{10} has investigated the notional indications from many astrophysical
procedures that establish the significance of natural dissimilarities in pressure. Herrera and Santos \cite {11}
determined the impacts of imperfect fluid in astral spherical objects for the behavior of leisurely gyration.
Webber \cite {12} examined the effects of anisotropy pressure in dense models with magnetic field.
Chakraborty et al. \cite {13} analyzed the gears of tangential pressure along with radial pressure in astrophysical
structures by quasi-spherical paradigms in the form of implosion. Garattini \cite {14} explored the configuration
of exposed peculiarity through mass limitation in metric $f(R)$ theory of gravity. Sharif and his traitors \cite {15a}-\cite{15g}
investigated the various fascinating consequence by the association of inhomogeneous energy density and Weyl tensor
in dense objects during the collapse in GR as well as in modified theories of gravity. Cognola et al.\cite{16} have introduced the results of the spherical imploding object
for black hole (BH) in $f(R)$ context by the parameter of positive Ricci invariant $R$. Capozziello et al. \cite{17} examined
the modified Lan$\acute{e}$-Emden equation by $N$ parameter in context of $f(R)$ and study the hydrostatic periods of astral
models. Copeland et al.\cite{18} have analyzed the several concepts to confer the expanding behavior of universe. Amendola et al.
\cite{19} discussed the feasible paradigms of metric $f(R)$ gravity in context of both Einstein and Jordan frames.

Chandrasekhar \cite{20} introduced the impacts of perfect fluid collapsing source and deliberate the constancy of dynamics.
Herrera et al. \cite{21} discussed the effects of dissipative variables on the gravitating imploding source and investigate
that heat dissipation makes the scalar structure constant. Abbas et al. \cite{22} presented the results of compact star models of
anisotropic fluid distribution with cylindrical symmetric static spacetime geometry. Mak and harko \cite{23} studied the kind of exact
results of the field equations by taking spherical structure. They also describe that energy density along with tangential
and radial pressure are limited and increase at the core of imperfect fluid object. Rahaman et al. \cite{24} prolonged the
Krori-Barua solution with spherical surface of static spacetime for the examination of charge imperfect fluid. Herrera and his collaborators \cite{25}-\cite{25t} analyzed the stability and dynamics of many gravitating source using various form of fluid. In few investigations they have used the non-casual as well as casual approaches to discuss the dynamics of gravitating source. Nolan \cite {46} discussed the gravitational collapse of counter gyrating dust cloud with cylindrical paradigms to find the naked peculiarity. Hayward \cite {47} produced the results of cylindrical geometry for black holes,
gravitational waves and cosmic strings. Chan \cite{26} has investigated the role of shear viscosity act in the imploding course.

In this paper, we analyze the dynamics of gravitational collapse with cylindrical symmetry in the context of $f(R)$ gravity. The
main persistence of this study is to examine the viscous dissipative collapse in cylindrical spacetime with full
causal approach.

The proposal of this work as follows: In section 2, we describe the cylindrical symmetric spacetime and isotropic collapsing
matter distribution in $f(R)$ theory of gravity. In section 3 Einstein modified field equations and
the dynamical equations have been acquired by gravitational cylindrical source. Section 4 is devoted to the transportation
equations attained in frame of $M\ddot{u}ller$-Israel-Stewart formalism \cite {48}-\cite {50}, then transportation equations
are associated by dynamical equations. Finally, the conclusions of the study have been discussed in the last section.

\section{Isotropic gravitational collapsing matter distribution fenced with cylindrical astral structure}

In general relativity (GR) Einstein-Hilbert (EH) action can be expressed as,

\begin{equation}\label{1}
S_{EH}=\frac{1}{2\kappa}\int d^4x\sqrt{-g}R.
\end{equation}
The prolonged formulation of Einstein-Hilbert (EH) in  $f(R)$ context is,
\begin{equation}\label{2}
S_{modif}=\frac{1}{2\kappa}\int d^4x\sqrt{-g}f(R),
\end{equation}
where $f(R)$ is arbitrary function of Ricci scalar $R$, $T_{\alpha\beta}$ is isotropic matter tensor and $\kappa$ is coupling
constant. The devising of equation (\ref{2}) in form of $f(R)$ theory have the following Einstein field equations,
\begin{equation}\label{3}
F(R)R_{\alpha\beta}-\frac{1}{2}f(R)g_{\alpha\beta}-\nabla_{\alpha}\nabla_{\beta}F(R)+g_{\alpha\beta}\nabla^\alpha\nabla_{\alpha}F(R)=\kappa T_{\alpha\beta},
\end{equation}
Here $F(R)=\frac{d f(R)}{d R}$, $\nabla^\alpha\nabla_{\alpha}$ is the D' Alembert operator and $\nabla_{\alpha}$ denotes
the covariant derivative. Above equations may be expressed in Einstein tensor as given below,
\begin{equation}\label{4}
G_{\alpha\beta}=\frac{\kappa}{F}\left(T_{\alpha\beta}^m+T_{\alpha\beta}^D\right),
\end{equation}
where
\begin{equation}\label{5}
T_{\alpha\beta}^D=\frac{1}{\kappa}\left[\frac{f(R)-RF(R)}{2}g_{\alpha\beta}+\nabla_{\alpha}\nabla_{\beta}F(R)-g_{\alpha\beta}\nabla^\alpha\nabla_{\alpha}F(R)\right],
\end{equation}
represent the stress action matter energy tensor.
The cylindrically symmetric non-static spacetime structure which contains the dissipative fluid defines interior metric given by,

\begin{equation}\label{6}
{ds{_-}}^{2}=-A^{2}dt^{2}+B^{2}dr^{2}+C^{2}
d\phi^{2}+D^{2}dz^{2},
\end{equation}
where $A=A(t,r)$, $B=B(t,r)$, $C=C(t,r)$, $D=D(t,r)$. We labeled the coordinates $x^{0}=t$, $x^{1}=r$, $x^{2}=\phi$ and $x^{3}=z$.
The exterior metric is given by,
\begin{equation}\label{ex}
ds^{2}_{+}=-\left(-\frac{2M(\nu)}{\tilde{R}}\right)d\nu^{2}-2d\nu d\tilde{R} + \tilde{R}^{2}(d\phi^{2}+\gamma^{2}dZ^2),
\end{equation}
where $M(\nu)$ is mass and $\gamma^{2}=-\frac{\Lambda}{3}$, $\Lambda$ is cosmological constant.
The tensor $T_{\alpha\beta}^-$ of the dissipative collapsing matter distribution is as,
\begin{eqnarray}\label{7}
&&T_{\alpha\beta}^-=(\mu+p+\Pi)V_{\alpha}V_{\beta}+(p+\Pi)g_{\alpha\beta}+q_{\alpha}V_{\beta}+q_{\beta}V_{\alpha}+\epsilon l_{\alpha} l_{\beta}+\pi_{\alpha\beta},\\\nonumber&&
\end{eqnarray}
Here $\mu$ represent energy density,  $p$ is the isotropic pressure, $\Pi$ the bulk viscosity, $q^{\alpha}$
the heat flux, $\pi_{\alpha\beta}$ is the shear viscosity, $\epsilon$ the radiation density, $V^{\alpha}$ the
four velocity of the fluid . Furthermore, $l^{\alpha}$ denotes radial null four vector. The above extents
gratify the results as below.
\begin{eqnarray}\label{8}
&&V^{\alpha}V_{\alpha}=-1, \quad V^{\alpha}q_{\alpha}=0, \quad l^{\alpha}V_{\alpha}=-1, \quad l^{\alpha}l_{\alpha}=0,\\
&&\pi_{\mu\nu}V^{\nu}=0, \quad \pi_{[\mu\nu]}=0, \quad  \pi_{\alpha}^\alpha=0.
\end{eqnarray}
In general non-reversible thermodynamics in \cite{51, 52}.
\begin{equation}\label{9}
\pi_{\alpha\beta}=-2\eta\sigma_{\alpha\beta},  \quad\quad    \Pi=-\zeta\Theta,
\end{equation}
wherever $\eta$ is the factor of shear viscosity and $\zeta$ denotes the factor of bulk viscosity, $\sigma_{\alpha\beta}$ and $\Theta$ are the shear tensor and heat flow expansion respectively.

The shear tensor $\sigma_{\alpha\beta}$ is defined by,
\begin{equation}\label{10}
\sigma_{\alpha\beta}=V_{(\alpha;\beta)}+a_{(\alpha V_{\beta})}-\frac{1}{3}\Theta h_{\alpha\beta},
\end{equation}
where the acceleration $a_{\alpha}$ and the expansion $\Theta$ are recognized by,
\begin{equation}\label{11}
a_{\alpha}=V_{\alpha;\beta}V^{\beta},\quad\quad \Theta=V_{;\alpha}^\alpha,
\end{equation}
and $h_{\alpha\beta}=g_{\alpha\beta}+V_{\alpha}V_{\beta}$ is the projector onto the hypersurface
orthogonal to the four velocity.
let us define the following quantities for the given metric,
\begin{equation}\label{12}
V^{\alpha}=A^{-1}\delta_{0}^\alpha,\quad\quad  q^{\alpha}=qB^{-1}\delta_{1}^\alpha, \quad\quad l^{\alpha}=A^{-1}\delta_{0}^\alpha+B^{-1}\delta_{1}^\alpha,
\end{equation}
where $q$ depending on $t$ and $r$. Also it trails from (\ref{8}) so that,
\begin{equation}\label{13}
\pi_{0\alpha}=0,  \pi_{1}^{1}= -2\pi_{2}^{2}=-2\pi_{3}^{3}.
\end{equation}
In a more explicit formation we can write,
\begin{equation}\label{pi}
\pi_{\alpha\beta}=\Omega\left(\chi_{\alpha}\chi_{\beta}-\frac{1}{3}h_{\alpha\beta}\right),
\end{equation}
Here $\chi^\alpha$ is a unit four vector along the radial direction, sustaining
\begin{equation}\label{14}
\chi^\alpha\chi_{\alpha}=1, \quad \chi^\alpha V_{\alpha}=0, \quad \chi^\alpha=B^{-1}\delta_{1}^\alpha,
\end{equation}
and $\Omega=\frac{3}{2}\pi_{1}^{1}$.
With Eq. (\ref{12}), we obtain the following non null components of shear tensor.
\begin{equation}\label{15}
\sigma_{11}=\frac{B^2}{3A}\left[\Sigma_{1}-\Sigma_{3}\right],  \sigma_{22}=\frac{C^2}{3A}\left[\Sigma_{2}-\Sigma_{1}\right],
\sigma_{33}=\frac{D^2}{3A}\left[\Sigma_{3}-\Sigma_{2}\right].
\end{equation}
Now
\begin{equation}\label{16}
\sigma_{\alpha\beta}\sigma^{\alpha\beta}=\sigma^2=\frac{1}{3A^2}\left [\Sigma_{1}^2+\Sigma_{2}^2+\Sigma_{3}^2\right],
\end{equation}
Here
\begin{equation}\label{17}
\Sigma_{1}=\frac{\dot{B}}{B}-\frac{\dot{C}}{C}, \Sigma_{2}=\frac{\dot{C}}{C}-\frac{\dot{D}}{D},
\Sigma_{3}=\frac{\dot{D}}{D}-\frac{\dot{B}}{B},
\end{equation}
and here dot denotes derivative with respect to $t$.

Using Eqs.(\ref{11}) and (\ref{12}), we get
\begin{equation}\label{18}
a_{1}=\frac{\acute A}{A}, \quad\Theta=\frac{1}{A}\left(\frac{\dot B}{B}+\frac{\dot C}{C}+\frac{\dot D}{D}\right),
\end{equation}
where prime represents derivative with respect to $r$ .
\section{ Dynamical Equations in $f(R)$ gravity}

The modified field Eqs. (\ref{4}) for cylindrical symmetric spacetime in $f(R)$ metric formalism gives the following system of equations.
\begin{eqnarray}\nonumber
&&\left(\frac{A}{B}\right)^2\left[-\frac{C''}{C}-\frac{D''}{D}+\frac{B'}{B}\left(\frac{C'}{C}+\frac{D'}{D}\right)-\frac{C'D'}{CD}\right]\\\nonumber&&+\left(\frac{\dot{B}\dot{C}}{BC}
+\frac{\dot{B}\dot{D}}{BD}+\frac{\dot{C}\dot{D}}{CD}\right)=\frac{\kappa}{F}\Big[\left(\mu+\epsilon\right)A^2
+\frac{A^2}{\kappa}\{-\left(\frac{f(R)-R F(R)}{2}\right)\\&&-\frac{\dot F}{A^2}\left(\frac{\dot B}{B}+\frac{\dot C}{C}+\frac{\dot D}{D}\right)+\frac{F''}{B^2}+\frac{F'}{B^2}\left(\frac{C'}{C}+\frac{D'}{D}-\frac{B'}{B}\right)\}\Big],\label{19}
\end{eqnarray}

\begin{equation}\label{h1}
-\left(\frac{\dot C'}{C}+\frac{\dot D'}{D}-\frac{\dot B C'}{BC}-\frac{\dot B D'}{BD}-\frac{\dot C A'}{AC}-\frac{\dot D A'}{AD}\right)= \frac{\kappa}{F}\left[-\left(q+\epsilon\right)AB+\frac{1}{\kappa}\left(\dot F'-\frac{A'\dot F}{A}-\frac{\dot B F'}{B}\right)\right],
\end{equation}

\begin{eqnarray}\nonumber
&&-\left(\frac{B}{A}\right)^2\left(\frac{\ddot C}{C}+\frac{\ddot D}{D}+\frac{\dot C\dot D}{CD}-\frac{\dot A\dot C}{AC}-\frac{\dot A\dot D}{AD}\right)+\left(\frac{C'D'}{CD}+\frac{A'C'}{AC}+\frac{A'D'}{AD}\right)\\\nonumber&&=\frac{\kappa}{F}\Big[\left(p+\Pi+\epsilon+2\frac{\Omega}{3}\right)B^2
+\frac{B^2}{\kappa}\Big\{\left(\frac{f(R)-R F(R)}{2}\right)+\frac{\ddot F}{A^2}\\&&+\frac{\dot F}{A^2}\left(\frac{\dot C}{C}+\frac{\dot D}{D}-\frac{\dot A}{A}\right)-\frac{F'}{B^2}\left(\frac{A'}{A}+\frac{C'}{C}+\frac{D'}{D}\right)\Big\}\Big],\label{20}
\end{eqnarray}

\begin{eqnarray}\nonumber
&&-\left(\frac{C}{A}\right)^2\left[\frac{\ddot B}{B}+\frac{\ddot D}{D}-\frac{\dot A}{A}\left(\frac{\dot B}{B}+\frac{\dot D}{D}\right)+\frac{\dot B\dot D}{BD}\right]+\left(\frac{C}{B}\right)^2 \Big[\frac{A''}{A}+\frac{D''}{D}-\frac{A'}{A}\left(\frac{B'}{B}-\frac{D'}{D}\right)\\\nonumber&&-\frac{B'D'}{BD}\Big]=\frac{\kappa}{F} \Big[\left(p+\Pi-\frac{\Omega}{3}\right)C^2
-\frac{C^2}{\kappa}\{-\left(\frac{f(R)-R F(R)}{2}\right)\\&&-\frac{\ddot F}{A^2}+\frac{F''}{B^2}+\frac{\dot F}{A^2}\left(\frac{\dot A}{A}
-\frac{\dot B}{B}-\frac{\dot C}{C}-\frac{\dot D}{D}\right)+\frac{F'}{B^2}\left(\frac{A'}{A}-\frac{B'}{B}+\frac{C'}{C}+\frac{D'}{D}\right)\}\Big],\label{21}
\end{eqnarray}
\begin{eqnarray}\nonumber
&&-\left(\frac{D}{A}\right)^2\left[\frac{\ddot B}{B}+\frac{\ddot C}{C}-\frac{\dot A}{A}\left(\frac{\dot B}{B}+\frac{\dot C}{C}\right)+\frac{\dot B\dot C}{BC}\right]\\\nonumber&&+\left(\frac{D}{B}\right)^2\Big[\frac{A''}{A}+\frac{C''}{C}-\frac{A'}{A}\left(\frac{B'}{B}-\frac{C'}{C}\right)-\frac{B'C'}{BC}\Big]=\frac{\kappa}{F}\Big[\left(p+\Pi-\frac{\Omega}{3}\right)D^2
\\\nonumber&&-\frac{D^2}{\kappa}\{-\left(\frac{f(R)-R F(R)}{2}\right)-\frac{\ddot F}{A^2}+\frac{F''}{B^2}+\frac{\dot F}{A^2}\left(\frac{\dot A}{A}
-\frac{\dot B}{B}-\frac{\dot C}{C}-\frac{\dot D}{D}\right)\\&&+\frac{F'}{B^2}\left(\frac{A'}{A}-\frac{B'}{B}+\frac{C'}{C}+\frac{D'}{D}\right)\}\Big].\label{22}
\end{eqnarray}
The energy of gravitating source per specific length defined by cylindrical symmetric space-time is given by \cite {47}, \cite {53}-\cite {55}.
\begin{equation}\label{23}
E=\frac{\left(1-l^{-2}\nabla^{a}r\nabla_{a}r\right)}{8}.
\end{equation}
For a cylindrical symmetric paradigms by killing vectors, the circumference radius $ \rho $ and specific
length $ l $ and moreover, arial radius $ r $ are described in \cite {47}, \cite {53}-\cite {55}.

$ \rho^2=\xi_{(1)a}\xi^a_{(1)},  l^2=\xi_{(2)a}\xi^a_{(2)}, $ so that $ r=\rho l $.

In the whole interior region C-energy has the following form \cite{15d}

\begin{equation}\label{24}
m(r,t)=El=\frac{l}{8}+\frac{1}{8D}\left[\frac{1}{A^2}\left(C\dot{D}+\dot{C}D\right){^2}-\frac{1}{B^2}\left(CD'+C'D\right){^2}\right].
\end{equation}

The proper time and radial derivatives are given by, through Misner and Sharp technique \cite{6}.
\begin{equation}\label{25}
D_{T}=\frac{1}{A}\frac{\partial}{\partial t}, D_{C}=\frac{1}{C'}\frac{\partial}{\partial r},
\end{equation}
Where $C$ is the areal radius of a spherical surface inside the limit. The
velocity of the collapsing matter is described by the proper time derivative of $C$ and $D$ \cite{15f},
i.e.,
\begin{equation}\label{26}
U=D_{T}C=\frac{\dot C}{A}, V=D_{T}D=\frac{\dot D}{A},
\end{equation}
which is always negative. Using this result, Eq. (\ref{24}), indicates that
\begin{equation}\label{27}
\tilde{E}\equiv\frac{C'}{B}=\left[\left(\frac{VC+UD}{D}\right)^2-\frac{8}{D}\left(m(r,t)-\frac{l}{8}\right)\right]^\frac{1}{2}-\frac{CD'}{BD}.
\end{equation}

The rate of change of mass with respect to proper time in equation (\ref{24}), with
the use of Eqs. (\ref{19}), (\ref{h1}), (\ref{20})-(\ref{22}) is given by
\begin{eqnarray}\nonumber
&&D_{T} m(r,t)=\frac{CD}{F}\Big[-4\pi\left\{\left(\mu+2\epsilon-p-\Pi+\frac{4\Omega}{3}\right)U+\tilde{E}(q+\epsilon)\right\}\\\nonumber&&
-U\Big\{\frac{\tilde{E}^2}{C'^2}\left(\frac{F'C'}{C}+\frac{F'D'}{D}-\frac{3\tilde{E}F'D_{C}B}{2}\right)+\frac{\tilde{E}^2}{2}\left(\frac{D_{C}AD_{T}F}{\dot{C}}+\frac{3F''}{C'^2}\right)\\\nonumber&&
+\frac{\tilde{E}^2F}{D}\left(D_{CC}D+\frac{(D_{C}D)^2}{4D}+\frac{D_{C}D}{4C}\right)+\frac{\tilde{E}^2FD_{CC}C}{C}-\frac{\tilde{E}^3FD_{C}B}{C'}\left(\frac{1}{C}+\frac{D_{C}D}{D}\right)\\\nonumber&&
-\frac{F}{B}\left(\frac{VD_{T}B}{D}+D_{TT}B\right)-\frac{F}{4}\left(\frac{D_{TT}C}{C}+\frac{D_{TT}D}{D}\right)+\frac{RF(R)}{2}-\frac{f(R)}{2}\Big\}\\\nonumber&&
-U^3\left\{\frac{\dot{F}}{\dot{C}}\left(\frac{D_{T}A}{2\dot{C}}-\frac{1}{C}\right)-\frac{\ddot{F}}{2\dot{C}^2}+\frac{FD_{T}A}{4C\dot{C}}\right\}\\\nonumber&&
+U^2\Big\{\frac{\tilde{E}^2}{\dot{C}}\left(\frac{3\dot F D_{T}B}{2\tilde{E}C'}-\frac{F'D_{C}A}{2C'}\right)+\frac{V\dot F}{\dot C D}+\frac{F D_{T}B}{B}\left(\frac{1}{C}-\frac{D_{T}A}{\dot C}\right)\\\nonumber&&-\frac{\tilde{E}^2F}{\dot C}\left(D_{CC}A-\frac{\tilde{E}D_{C}A D_{C}B}{C'}+\frac{D_{C}A}{2C}\right)+\frac{VF}{2D}\left(\frac{3}{4C}-\frac{D_{T}A}{\dot C}\right)\Big\}\\\nonumber&&
-\frac{V^2F}{2D}\left(\frac{\tilde{E}^2D_{C}A}{\dot D}-\frac{U}{2D}\right)-\frac{V^3F}{4D^2}\left(\frac{C}{D}+\frac{CD_{T}A}{\dot D}\right)\\\nonumber&&+\frac{VF}{4D}\left\{\frac{CD_{TT}D}{D}+D_{TT}C+\tilde{E}^2\left(\frac{C(D_{C}D)^2}{2D^2}-\frac{1}{2C}\right)\right\}\\\nonumber&&
-\frac{\tilde{E}^2}{2C'}\Big\{\tilde{E}D_{T}B\left(D_{C}F+\frac{F}{2C}-\frac{C(D_{C}D)^2F}{2D^2}\right)-F\left(\frac{D_{T}C'}{2C}+\frac{D_{T}D'}{2D}\right)\\&&
+\frac{FD_{C}D}{2D}\left(\frac{CD_{T}D'}{D}+D_{T}C'\right)-D_{T}F'\Big\}\Big].\label{28}
\end{eqnarray}
The overhead formation chiefs us to the distinction rate of mass energy in the cylinder having
radius C and the R.H.S of this equation pronounces the raising in energy in the inside boundary
of radius C. The value $(\mu+2\epsilon-p-\Pi+\frac{4\Omega}{3})$ is the effective radial pressure
whereas $\mu, \epsilon $ and $ \Pi $ are the energy density, the radiation pressure and the bulk
viscosity. The usual thermodynamic result is $\pi_{\alpha\beta}$ in the relaxation period and the
second value of the right-hand side illustrates the matter energy of the fluid, which is releasing
the cylindrical exterior. Moreover, all extra terms reflects the behavior of the dark source of the
dissipative matter under the formation of modified theory of gravity and with its differentiation, lessening the pressure
in the core of star due to continuous collapsing phase of the dark energy. Similarly, we can calculate
\begin{eqnarray}\nonumber
&&D_{C} m(r,t)=\frac{CD}{2F}\Big[\kappa\left\{\mu+2\epsilon+p+\Pi+\frac{2\Omega}{3}+\frac{U}{\tilde{E}}\left(q+\epsilon\right)\right\}\\\nonumber&&+\frac{U^3FD_{C}A}{2C\dot C}+U^2\left\{\frac{D_{T}FD_{C}A}{\dot C}+\frac{F}{C}\left(\frac{D_{C}D}{4D}-\frac{D_{T}A}{\dot C}\right)\right\}
-U\Big\{\tilde{E}D_{C}F\left(\frac{\tilde{E}D_{C}A}{\dot C}-\frac{D_{T}B}{C'}\right)\\\nonumber&&+\frac{D_{T}FD_{T}A}{\dot C}+F\left(\frac{1}{C'}\left(\frac{D_{T}C'}{2C}+\frac{D_{T}D'}{2D}-\frac{\tilde{E}D_{T}BD_{C}D}{D}\right)+\frac{\tilde{E}^2}{\dot C}\left(\frac{D_{C}A}{C}+\frac{D_{C}AD_{C}D}{D}\right)\right)\Big\}\\\nonumber&&-\frac{V^2F}{2D^2}\left\{C\left(\frac{U D_{C}A}{\dot C}+\frac{D_{C}D}{2D}\right)-1\right\}+\frac{VF}{2C'D}\left(\frac{CD_{T}D'}{D}+D_{T}C'\right)+\frac{U V F}{D}\left(\frac{1}{2C}-\frac{D_{T}A}{\dot C}\right)\\\nonumber&&-\tilde{E}^2F\Big\{\frac{D_{C}D}{2D^2}\left(CD_{CC}D+D_{C}D-\frac{C(D_{C}D)^2}{2D}-\frac{\tilde{E}CD_{C}BD_{C}D}{C'}\right)\\\nonumber&&
-\frac{1}{2C}\left(D_{CC}C-\frac{3D_{C}D}{2D}-\frac{\tilde{E}D_{C}B}{C'}\right)+\frac{1}{2D}\left(D_{C}D D_{CC}C-D_{CC}D\right)\Big\}\\\nonumber&&
-\frac{\tilde{E}}{C'}\left(D_{T}F D_{T}B+\tilde{E}^2D_{C}F D_{C}B+\frac{FD_{T}B D_{T}D}{D}\right)+D_{TT}F+\frac{F''\tilde{E}^2}{C'^2}\\&&-\frac{U D_{T}F'}{C'}+\frac{FD_{TT}C}{C}+\frac{FD_{TT}D}{D}\Big].\label{29}
\end{eqnarray}
This solution explains how variational quantities effect the matter distribution between the neighboring exteriors in
the object of radius C. The first two quantities on R.H.S and their picture relates with the overhead result expect for
the factor $p+\Pi+\frac{2\Omega}{3}$. The presence of this issue is due to the complex field equations. The outstanding
values signify input of dark energy (DE) because of the curvature matter. Applying integration  of Eq.(\ref{29}) with $C$, we establish
\begin{eqnarray}\nonumber
&&m(r,t)=\frac{1}{2}{\int_0^C}\frac{CD}{2F}\Big[\kappa\left\{\mu+2\epsilon+p+\Pi+\frac{2\Omega}{3}+\frac{U}{\tilde{E}}\left(q+\epsilon\right)\right\}\\\nonumber&&+\frac{U^3FD_{C}A}{2C\dot C}+U^2\left\{\frac{D_{T}FD_{C}A}{\dot C}+\frac{F}{C}\left(\frac{D_{C}D}{4D}-\frac{D_{T}A}{\dot C}\right)\right\}
-U\Big\{\tilde{E}D_{C}F\left(\frac{\tilde{E}D_{C}A}{\dot C}-\frac{D_{T}B}{C'}\right)\\\nonumber&&+\frac{D_{T}FD_{T}A}{\dot C}+F\left(\frac{1}{C'}\left(\frac{D_{T}C'}{2C}+\frac{D_{T}D'}{2D}-\frac{\tilde{E}D_{T}BD_{C}D}{D}\right)+\frac{\tilde{E}^2}{\dot C}\left(\frac{D_{C}A}{C}+\frac{D_{C}AD_{C}D}{D}\right)\right)\Big\}\\\nonumber&&-\frac{V^2F}{2D^2}\left\{C\left(\frac{U D_{C}A}{\dot C}+\frac{D_{C}D}{2D}\right)-1\right\}+\frac{VF}{2C'D}\left(\frac{CD_{T}D'}{D}+D_{T}C'\right)+\frac{U V F}{D}\left(\frac{1}{2C}-\frac{D_{T}A}{\dot C}\right)\\\nonumber&&-\tilde{E}^2F\Big\{\frac{D_{C}D}{2D^2}\left(CD_{CC}D+D_{C}D-\frac{C(D_{C}D)^2}{2D}-\frac{\tilde{E}CD_{C}BD_{C}D}{C'}\right)\\\nonumber&&
-\frac{1}{2C}\left(D_{CC}C-\frac{3D_{C}D}{2D}-\frac{\tilde{E}D_{C}B}{C'}\right)+\frac{1}{2D}\left(D_{C}D D_{CC}C-D_{CC}D\right)\Big\}\\\nonumber&&
-\frac{\tilde{E}}{C'}\left(D_{T}F D_{T}B+\tilde{E}^2D_{C}F D_{C}B+\frac{FD_{T}B D_{T}D}{D}\right)+D_{TT}F+\frac{F''\tilde{E}^2}{C'^2}\\&&-\frac{U D_{T}F'}{C'}+\frac{FD_{TT}C}{C}+\frac{FD_{TT}D}{D}\Big]dC.\label{30}
\end{eqnarray}
The above equation gives the total mass in terms of C-energy inside the cylinder with the contribution of $f(R)$ dark source term.
To perceive the dynamical action by assuming Misner and Sharp \cite {6,56} concept,
the contracted Bianchi identities are given as,
\begin{equation}\label{31}
\left(T^{(m)}_{\alpha\beta}+T^{(D)}_{\alpha\beta}\right)_{;\beta}V_{\alpha}=0,\quad\quad (T^{(m)}_{\alpha\beta}+T^{(D)}_{\alpha\beta})_{;\beta}\chi_{\alpha}=0,
\end{equation}
which produce
\begin{eqnarray}\nonumber
&&\frac{1}{A}\left[(\mu+\epsilon\dot{)}+\left(\mu+p+\Pi+2\epsilon+\frac{2\Omega}{3}\right)\frac{\dot B}{B}+\left(\mu+p+\Pi+\epsilon-\frac {\Omega}{3}\right)\left(\frac{\dot C}{C}+\frac{\dot D}{D}\right)\right]\\&&+
\frac{1}{B}\left[\left(q+\epsilon\right)'+\left(q+\epsilon\right)\left(\frac{2A'}{A}+\frac{(CD)'}{CD}\right)\right]-D_{1}=0.\label{32}
\end{eqnarray}
\begin{eqnarray}\nonumber
&&\frac{1}{A}\left[ (q+\epsilon\dot{)}+(q+\epsilon)\left(2\frac{\dot B}{B}+\frac{\dot C}{C}+\frac{\dot D}{D}\right)\right]\\\nonumber&&\label{33}+\frac{1}{B}\left[\left(p+\Pi+\epsilon+\frac{2\Omega}{3}\right)'+\left(\mu+p+\Pi+2\epsilon+\frac{2\Omega}{3}\right)\frac{A'}{A}
+\left(\epsilon+\Omega\right)\frac{ (CD)'}{CD}\right]+D_{2}=0,\\
\end{eqnarray}
Here $D_{1}$ and $D_{2}$ are the expressions of dark energy having in Appendix of equations (\ref{52}) and (\ref{53})  respectively.

Using Eqs. (\ref{20}),(\ref{25})-(\ref{27}), it follows that
\begin{eqnarray}\nonumber
&&D_{T}(\mu+\epsilon)+\frac{1}{3}\left(3\mu+3p+3\Pi+4\epsilon\right)\Theta+\frac{1}{3}\left(\epsilon+\Omega\right)\left(\frac{2\dot B}{A B}-\frac{U}{C}-\frac{V}{D}\right)\\&&+\tilde{E}D_{C}(q+\epsilon)+(q+\epsilon)\left(2\frac{a_{1}}{B}+\frac{\tilde{E}}{C}+\frac{D'}{BD}\right)-D_{1}=0.\label{34}
\end{eqnarray}
\begin{eqnarray}\nonumber
&&D_{T}(q+\epsilon)+(q+\epsilon)\left(\Theta+\frac{D_{T}B}{B}\right)+\tilde{E}D_{C}\left(p+\Pi+\epsilon+\frac{2\Omega}{3}\right)\\&&+
\left(\mu+p+\Pi+2\epsilon+\frac{2\Omega}{3}\right)\frac{a_{1}}{B}+\left(\epsilon+\Omega\right)\frac{(CD)'}{B C D}+D_{2}=0.\label{35}
\end{eqnarray}
The acceleration $D_{T}U$ of the dissipative viscous collapsing source is
achieved by using Eqs. (\ref{20}),(\ref{25})-(\ref{27}), are
\begin{eqnarray}\nonumber
&&D_{T}U=-\frac{m(r,t)}{CD}+\frac{l}{8CD}-\frac{4\pi C}{F}\left(p+\Pi+\epsilon+\frac{2\Omega}{3}\right)+\left(\frac{\tilde{E}}{2}+\frac{D'C}{2BD}\right)\frac{a_{1}}{B}
\\\nonumber&&-\frac{C}{F}\left[\frac{D_{TT}F}{2}+\frac{U D_{T}F}{2}\left(\frac{1}{C}-\frac{D_{T}A}{\dot C}\right)+\frac{V D_{T}F}{2D}
-\frac{\tilde{E} D' D_{C}F}{2BD}-\frac{\tilde{E}^2 D_{C}F}{2}\left(\frac{U D_{C}A}{\dot C}+\frac{1}{C}\right)\right]\\\nonumber&&+\frac{D_{TT}C}{2}-\frac{CD_{TT}D}{2D}-\frac{D_{T}A}{2A}\left(U-\frac{CV}{D}\right)
-\frac{1}{4D}\left(UV-\frac{\tilde{E}D'}{B}\right)+\frac{1}{8D^2}\left(CV^2-\frac{CD'^2}{B^2}\right)\\\label{36}&&+\frac{1}{8C}\left(U^2-\tilde{E}^2\right)-\frac{C(f(R)-R F(R))}{4F}.
\end{eqnarray}
After, inserting $\frac{a_{1}}{B}$ from (\ref{36}) into (\ref{35}),we get
\begin{eqnarray}\nonumber
&&\left(\mu+p+\Pi+2\epsilon+\frac{2\Omega}{3}\right)D_{T}U\\\nonumber&&=-\left(\mu+p+\Pi+2\epsilon+\frac{2\Omega}{3}\right)\Big[\frac{m(r,t)}{CD}-\frac{l}{8CD}
+\frac{4\pi C}{F}\left(p+\Pi+\epsilon+\frac{2\Omega}{3}\right)\\\nonumber&&+\frac{C(f(R)-R F(R))}{4F}+\frac{C D_{TT}F}{2F}+\left(\frac{1}{C}-\frac{D_{T}A}{\dot C}\right)\frac{U C D_{T}F}{2F}+\frac{VC D_{T}F}{2DF}-\frac{\tilde{E}CD'D_{C}F}{2BDF}\\\nonumber&&-\frac{\tilde{E}^2C D_{C}F}{2F}\left(\frac{1}{C}+\frac{U D_{C}A}{\dot C}\right)-\frac{D_{TT}C}{2}+\frac{C D_{TT}D}{2D}+\frac{D_{T}A}{2A}\left(U-\frac{VC}{D}\right)\\\nonumber&&+
\frac{1}{4D}\left(UV-\frac{\tilde{E}D'}{B}\right)-\frac{1}{8D^2}\left(V^2C-\frac{CD'^2}{B^2}\right)-\frac{1}{8C}\left(U^2-\tilde{E}^2\right)\Big]\\\nonumber&&
-\left(\frac{\tilde{E}}{2}+\frac{D'C}{2BD}\right)\left[\tilde{E}D_{C}\left(p+\Pi+\epsilon+\frac{2\Omega}{3}\right)+\frac{\tilde{E}}{C}\left(\epsilon+\Omega\right)
+\frac{D'}{BD}\left(\epsilon+\Omega\right)\right]\\&&\label{37}
-\left(\frac{\tilde{E}}{2}+\frac{D'C}{2BD}\right)\left[D_{T}(q+\epsilon)+(q+\epsilon)\left(\Theta+\frac{D_{T}B}{B}\right)+D_{2}\right].
\end{eqnarray}
Now, to evaluate the term $\left(\mu+p+\Pi+2\epsilon+\frac{2\Omega}{3}\right)$ seems on the L.H.S and
as well as comes on R.H.S, this is the effective inertial mass, and rendering to the equivalence principle
it is too recognized as passive gravitational mass. On the R.H.S, the first term of square bracket quantity
explicates the impacts of collapsing variables on the active gravitational mass of the cylindrical collapsing
object in dark source with f(R) metric theory, this datum has been pointed out early by Herrera et al. \cite{25g} in the context of GR.
Now in the second square bracket there are the gradient of the whole active pressure which is prejudiced with
collapsing variables, radiating density. The final bracket comprises unalike additions because of the collapsing
matter nature of the structure. The second value in this bracket is positive inferring that outgoing of $q>0$ and
$\epsilon>0$ decreases the total energy of diminishing matter, which reduces the degree of collapse. The last term
of this bracket describes the dark energy source of the dissipative gravitating collapse.
\section{Transport Equations}
The purpose of this section is to confer a full causal technique for the viscid dissipative
gravitating collapse of astral paradigms accompanied by heat transference. This infers that all
collapsing variables should please the transportation equations attained from causal thermodynamics.
Therefore, we utilize the transportation equations for heat, bulk and shear viscosity from
$M\ddot{u}ller$-Israel-Stewart formalism \cite {48}-\cite {50} for dissipative source. Herrera et al.
\cite{25g} discussed transportation equations for heat, bulk and shear viscosity. The entropy flux is given by
\begin{eqnarray}\nonumber
S^{\mu}=S n V^{\mu}+\frac{q^{\mu}}{T}-(\beta_{0}\Pi^{2}+\beta_{1}q_{\nu}q^{\nu}+\beta_{2}\pi_{\nu\kappa}\pi^{\nu\kappa})\frac{V^{\mu}}{2T}+\frac{\alpha_{0}\Pi q^{\mu}}{T}+\frac{\alpha_{1}\pi^{\mu\nu} q_{\nu}}{T},\\\label{38}
\end{eqnarray}
where $ \beta_{1}$ and $\beta_{2}$ are thermodynamic factors for unalike additions to entropy density,
$\alpha_{0}$ and $\alpha_{1}$ are thermodynamics heat coupling factors, $T$ is temperature.

Furthermore, from the Gibbs equation and Bianchi identities, it follows that
\begin{eqnarray}\nonumber
&&T S_{;\alpha}^{\alpha}=-\Pi\left[V_{;\alpha}^{\alpha}-\alpha_{0}q_{;\alpha}^{\alpha}+\beta_{0}\Pi_{;\alpha}V^{\alpha}+\frac{T}{2}\left(\frac{\beta_{0}}{T}V^{\alpha}\right)_{;\alpha}\Pi\right]\\\nonumber&&
-q^{\alpha}[h_{\alpha}^{\mu}(ln T)_{,\mu}(1+\alpha_{0}\Pi)+V_{\alpha;\mu}V^{\mu}-\alpha_{0}\Pi_{;\alpha}\\\nonumber&&
-\alpha_{1}\pi_{\alpha;\mu}^{\mu}+\alpha_{1}\pi_{\alpha}^{\mu}h_{\mu}^{\beta}(ln T)_{,\beta}+\beta_{1}q_{\alpha;\mu}V^{\mu}+\frac{T}{2}\left(\frac{\beta_{1}}{T}V^{\mu}\right)_{;\mu}q_{\alpha}]\\&&\label{39}
-\pi^{\alpha\mu}\left[\sigma_{\alpha\mu}-\alpha_{1}q_{\mu;\alpha}+\beta_{2}\pi_{\alpha\mu;\nu}V^{\nu}+\frac{T}{2}\left(\frac{\beta_{2}}{T}V^{\nu}\right)_{;\nu}\pi_{\alpha\mu}\right].
\end{eqnarray}
Finally, by the standard procedure, the constitutive transport equations follow from the requirement $S_{;\alpha}^\alpha\geq0$
\begin{eqnarray}\label{40}
\tau_{0}\Pi_{,\alpha}V^\alpha+\Pi=-\zeta\theta+\alpha_{0}\zeta q_{;\alpha}^\alpha-\frac{1}{2}\zeta T\left(\frac{\tau_{0}}{\zeta T}V^\alpha\right)_{;\alpha}\Pi.
\end{eqnarray}
\begin{eqnarray}\nonumber
\tau_{1}h_{\alpha}^\beta q_{\beta;\mu}V^\mu+q_{\alpha}&&=-\kappa\left[h_{\alpha}^\beta T_{,\beta}(1+\alpha_{0}\Pi)+\alpha_{1}\pi_{\alpha}^\mu h_{\mu}^\beta T_{,\beta}+T(a_{\alpha}-\alpha_{0} \Pi_{;\alpha}-\alpha_{1}\pi_{\alpha;\mu}^\mu)\right]\\&&-\frac{1}{2}\kappa T^2\left(\frac{\tau_{1}}{\kappa T^2}V^\beta\right)_{;\beta}q_{\alpha}\\
\tau_{2}h_{\alpha}^\mu h_{\beta}^\nu \pi_{\mu\nu;\rho}V^\rho+\pi_{\alpha\beta}&&=-2\eta\sigma_{\alpha\beta}+2\eta \alpha_{1}q_{<\beta;\alpha>}-\eta T\left(\frac{\tau_{2}}{2\eta T}V^\nu\right)_{;\nu}\pi_{\alpha\beta}\label{41},
\end{eqnarray}
with
\begin{eqnarray}\label{42}
q_{<\beta;\alpha>}=h_{\beta}^\mu h_{\alpha}^\nu\left(\frac{1}{2}(q_{\mu;\nu}+q_{\nu;\mu})-\frac{1}{3}q_{\sigma;\kappa}h^{\sigma\kappa} h_{\mu\nu}\right),
\end{eqnarray}
and where the relaxational times are given by
\begin{eqnarray}\label{43}
\tau_{0}=\zeta\beta_{0},\quad\quad \tau_{1}=\kappa T \beta_{1},\quad\tau_{2}=2 \eta \beta_{2},
\end{eqnarray}
Here, $\zeta$ and $\eta$ are the factors of bulk and shear viscosity. Using the interior metric
of cylindrical structure into eqs. (\ref{40})-(\ref{41}) have following set of equations,
\begin{eqnarray}\nonumber
&&\tau_{0}\dot \Pi=-\left(\zeta+\frac{\tau_{0}}{2}\Pi\right)A\Theta+\frac{A}{B}\alpha_{0}\zeta\left[q'+q\left(\frac{A'}{A}+\frac{C'}{C}+\frac{D'}{D}\right)\right]\\&&\label{44}-\Pi\left[\frac{\zeta T}{2}(\frac{\tau_{0}}{\zeta T}\dot{)}+A\right],\\\nonumber
&&\tau_{1}\dot q=-\frac{A}{B}\kappa\Big\{T'\left(1+\alpha_{0}\Pi+\frac{2}{3}\alpha_{1}\Omega\right)+T\Big[\frac{A'}{A}-\alpha_{0}\Pi'\\\nonumber&&\label{45}-
\frac{2}{3}\alpha_{1}\left(\Omega'+\frac{A'}{A}\Omega+\frac{3}{2}\left(\frac{C'}{C}+\frac{D'}{D}\right)\Omega\right)\Big]\Big\}\\&&-q\left[\frac{\kappa T^2}{2} (\frac{\tau_{1}}{\kappa T^2}\dot{)}+\frac{\tau_{1}}{2} A \theta+A\right],\\\nonumber
&&\tau_{2}\dot \Omega=-\eta\left(\frac{2\dot B}{B}-\frac{\dot C}{C}-\frac{\dot D}{D}\right)+2\eta\alpha_{1}\frac{A}{B}q'-\Omega\left[\eta T (\frac{\tau_{2}}{2\eta T}\dot {)}+\frac{\tau_{2}}{2}A \Theta+A\right],\\\label{46}
\end{eqnarray}
Here, we have to analyze the effect of several dissipative variables on the cylindrical interior surface.
For this persistence, we use Eq.(\ref{45}) into Eq.(\ref{37}) and get
\begin{eqnarray}\nonumber
&&\left(\mu+p+\Pi+2\epsilon+\frac{2}{3}\Omega\right)(1-\Lambda)D_{T}U\\\nonumber&&=(1-\Lambda)F_{grav}+F_{hyd}+\tilde{E}\left(\frac{\tilde{E}}{2}+
\frac{D'C}{2BD}\right)\frac{\kappa}{\tau_{1}}\Big\{D_{C}T\left(1+\alpha_{0}\Pi+\frac{2}{3}\alpha_{1}\Omega\right)\\\nonumber&&-T\left[\alpha_{0}D_{C}\Pi+
\frac{2}{3}\alpha_{1}\left(D_{C}\Omega+\frac{3(CD)'}{2CDC'}\Omega\right)\right]\Big\}\\\nonumber&&+\left(\frac{\tilde{E}}{2}+
\frac{D'C}{2BD}\right)\left[\frac{\kappa T^2 q}{2\tau_{1}}D_{T}\left(\frac{\tau_{1}}{\kappa T^2}\right)-D_{T}\epsilon\right]\\&&-\left(\frac{\tilde{E}}{2}+
\frac{D'C}{2BD}\right)\left[\left(\frac{q}{2}+2\epsilon\right)\theta-\frac{q}{\tau_{1}}+(q+\epsilon)\frac{D_{T}B}{B}+D_{2}\right],\label{47}
\end{eqnarray}
where $F_{grav}$ and $F_{hyd}$ are representing as
\begin{eqnarray}\nonumber
&&F_{grav}=-\left(\mu+p+\Pi+2\epsilon+\frac{2\Omega}{3}\right)\Big[m(r,t)-\frac{l}{8}
+4\pi\left(p+\Pi+\epsilon+\frac{2\Omega}{3}\right)\frac{C^2D}{F}\\\nonumber&&+\frac{C^2D(f(R)-R F(R))}{4F}+\frac{C^2D D_{TT}F}{2F}+\frac{U C^2D D_{T}F}{2F}\left(\frac{1}{C}-\frac{D_{T}A}{\dot C}\right)+\frac{VC^2 D_{T}F}{2F}\\\nonumber&&-\frac{\tilde{E}C^2D'D_{C}F}{2BF}-\frac{\tilde{E}^2C^2D D_{C}F}{2F}\left(\frac{1}{C}+\frac{U D_{C}A}{\dot C}\right)-\frac{CD D_{TT}C}{2}+\frac{C^2 D_{TT}D}{2}\\\nonumber&&+\frac{CD D_{T}A}{2A}\left(U-\frac{VC}{D}\right)
+\frac{C}{4}\left(UV-\frac{\tilde{E}D'}{B}\right)-\frac{C}{8D}\left(V^2C-\frac{CD'^2}{B^2}\right)\\&&+-\frac{D}{8}\left(U^2-\tilde{E}^2\right)\Big]\frac{1}{CD},\\\nonumber
&&F_{hyd}=-\left(\frac{\tilde{E}}{2}+\frac{D'C}{2BD}\right)\Big[\tilde{E}D_{C}\left(p+\Pi+\epsilon+\frac{2\Omega}{3}\right)
+\frac{\tilde{E}}{C}\left(\epsilon+\Omega\right)\\\nonumber&&\label{48}+\frac{D'}{BD}\left(\epsilon+\Omega\right)\Big],
\end{eqnarray}
and $\Lambda$ is defined as
\begin{eqnarray}\label{49}
\Lambda=\frac{\kappa T}{\tau_{1}}\left(\mu+p+\Pi+2\epsilon+\frac{2}{3}\Omega\right)^{-1} \left(1-\frac{2}{3}\alpha_{1}\Omega\right).
\end{eqnarray}
 We use Eq.(\ref{44}) into Eq.(\ref{47}) and subsequently get
\begin{eqnarray}\nonumber
&&\left(\mu+p+\Pi+2\epsilon+\frac{2}{3}\Omega\right)(1-\Lambda+\Delta)D_{T}U=(1-\Lambda+\Delta)F_{grav}+F_{hyd}\\\nonumber&&+\frac{\tilde{E}\kappa}{\tau_{1}}\left(\frac{\tilde{E}}{2}+
\frac{D'C}{2BD}\right)\left\{D_{C}T\left(1+\alpha_{0}\Pi+\frac{2}{3}\alpha_{1}\Omega\right)-T\left[\alpha_{0}D_{C}\Pi+
\frac{2}{3}\alpha_{1}\left(D_{C}\Omega+\frac{3(CD)'}{2CD C'}\Omega\right)\right]\right\}\\\nonumber&&
-\tilde{E}\left(\frac{\tilde{E}}{2}+\frac{D'C}{2BD}\right)\left(\mu+p+\Pi+2\epsilon+\frac{2}{3}\Omega\right)\Delta \left(\frac{D_{C}q}{q}+\frac{(CD)'}{CD C'}\right)\\\nonumber&&+\left(\frac{\tilde{E}}{2}+\frac{D'C}{2BD}\right)\left[\frac{\kappa T^2 q}{2\tau_{1}}D_{T}\left(\frac{\tau_{1}}{\kappa T^2}\right)-D_{T}\epsilon\right]+\left(\frac{\tilde{E}}{2}+\frac{D'C}{2BD}\right)\left[\frac{q}{\tau_{1}}-(q+\epsilon)\frac{D_{T}B}{B}-D_{2}\right]\\\nonumber&&
+\left(\frac{\tilde{E}}{2}+\frac{D'C}{2BD}\right)\frac{\Delta}{\alpha_{0}\zeta q}\left(\mu+p+\Pi+2\epsilon+\frac{2}{3}\Omega\right)\left\{\left[1+\frac{\zeta T}{2}D_{T}(\frac{\tau_{0}}{\zeta T})\right]\Pi+\tau_{0}D_{T}\Pi\right\},\\\label{50}
\end{eqnarray}
where $\Delta$ is defined as
\begin{equation}\label{51}
\Delta=\alpha_{0}\zeta q \left(\mu+p+\Pi+2\epsilon+\frac{2}{3}\Omega\right)^{-1}\left(\frac{q+2\epsilon}{2\zeta+\tau_{0}\Pi}\right).
\end{equation}
From now, by captivating made about interpretation the causal transport equations and their association by the
dynamical equation, we trace that the term $1-\Lambda+\Delta$ distresses extensively the inner energy and
effective inertial mass density. This consequence is covenant with the finding of Herrera et al.\cite{25g}.
\section{Conclusions}
The maiden of the 20th century Weyl \cite{57} and Levi-Civita \cite{58} ongoing that work
with such static cylindrical symmetric structures after Einstein's notion of gravity. Later, the
ample work of spherical surfaces, hypothetical astronomer were observed to reconnoiter the effects
of astral dense objects that have spherical symmetry. The cosmological source including heat dissipation
and viscosity are very noticeable for searching the progression of dens objects. Therefore, it is
significant to involve the cylindrical symmetry in gravitating implosion.

In this study, we have arranged the dynamical equations which play meaningful role in progressively stages of a viscid dissipative gravitational cylindrical region. In way to perceive
the impacts of $f(R)$ terms on the dynamical progression of a gravitational stuff, we have alleged the
expedient formation of the collapsing variables which please the transport equations of heat radiation,
bulk and shear viscosity ensuing from causal thermodynamical approach. Moreover, the collapsing factors due
to viscosity and heat flux have been involved in the examination of dynamical equations. For extensive point
of view, we are mainly concerned in relaxation phase whose direction might be lesser or equal to
radiating time. Throughout the work of transportation equations for collapsing variables, we have
chosen to take the hyperbolic concept of collapse due to this concept is much consistent and has rare
problems than parabolic concept \cite {59}-\cite {61}.

A full causal technique has been implemented in \cite {25g} to evaluate the impacts of collapsing
variables on the spharically symmetric collapse: this work gives the evocative solutions which have
noteworthy consequences in astrophysics. The application of these solutions to some astral objects
infers that in a sidereal existence, the radiating transformation of the collapsing material might be
huge to yield an noticeable lessening in the gravitating action of the structure. It is pretty related to allusion
that thermodynamical viscid/heat radiating coupling factors have been kept as non-terminating due to
this theory gives the momentous foundation for the designing of variational astral model. Current scenario
\cite {62} took a partial technique to confer the impacts of bulk dissipative gravitating collapse
source in cylindrical surface of the star with deserting the thermodynamical heat radiating coupling
factors in the transport equations.

Using full causal technique the importance of dynamical viscous gravitational collapse
is to be seen that with dynamical equation (\ref{50}), which explicates how the measure of passive
gravitational is effected by the collapsing variables and thermodynamical heat radiating coupling
factors. In astral objects, happening all collapsing variables have much impacts, for instance
a galactic estimation of heat coupling factor $\kappa$ can present a rushing lessening in gravitational
force, thus ensuing in the setback of implosion \cite {25g}. Herrera at al. \cite {63} investigated the
bouncing action envisaging in mathematical paradigms in the current arithmetic calculation. They have also,  The dissipative collapse due to force of gravity and we ratify
that thermodynamical heat radiating coupling factors must be included as preceding in transport equations. The present work is utilized in future with other modified theories such as Gauss-Bonet and $f(R,T)$ framework. This work has been done for spherical symmetry with and without electromagnetic field \cite{h1}.
\section{Appendix}
\begin{eqnarray}\nonumber
&&D_{1}=\frac{A}{\kappa}\Big[\left\{\frac{1}{A^{2} B^{2}}\left(\dot F'-\frac{A'\dot F}{A}-\frac{\dot B F'}{B}\right)\right\}_{,1}\\\nonumber&&+
\left\{\frac{f(R)-R F(R)}{2A^{2}}-\frac{F''}{A^{2} B^{2}}+\frac{\dot F}{A^4}\left(\frac{\dot B}{B}+\frac{\dot C}{C}+\frac{\dot D}{D}\right)-\frac{F'}{A^{2} B^{2}}\left(\frac{C'}{C}+\frac{D'}{D}-\frac{B'}{B}\right)\right\}_{,0}\\\nonumber&&+\frac{2\dot A}{A^3}\left\{\frac{f(R)-R F(R)}{2}-\frac{F''}{B^{2}}+\frac{\dot F}{A^2}\left(\frac{\dot B}{B}+\frac{\dot C}{C}+\frac{\dot D}{D}\right)-\frac{F'}{B^{2}}\left(\frac{C'}{C}+\frac{D'}{D}-\frac{B'}{B}\right)\right\}\\\nonumber&&
+\frac{\dot B}{A^{2} B}\left\{-\frac{\ddot F}{A^{2}}-\frac{F''}{B^{2}}+\frac{\dot F}{A^{2}}\left(\frac{\dot A}{A}+\frac{\dot B}{B}\right)+\frac{F'}{B^{2}}\left(\frac{A'}{A}+\frac{B'}{B}\right)\right\}\\\nonumber&&+\frac{\dot C}{A^{2} C}\left\{-\frac{\ddot F}{A^{2}}+\frac{\dot F}{A^{2}}\frac{\dot A}{A}+\frac{F'}{B^{2}}\frac{A'}{A}\right\}+\frac{\dot D}{A^{2} D}\left\{-\frac{\ddot F}{A^{2}}+\frac{\dot F}{A^{2}}\frac{\dot A}{A}+\frac{F'}{B^{2}}\frac{A'}{A}\right\}\\\nonumber&&
+\frac{1}{A^{2} B^{2}}\left(\dot F'-\frac{A'\dot F}{A}-\frac{\dot B F'}{B}\right)\left(\frac{3A'}{A}+\frac{B'}{B}+\frac{C'}{C}+\frac{D'}{D}\right)\Big].\\\label{52}
\end{eqnarray}
\begin{eqnarray}\nonumber
&&D_{2}=\frac{B}{\kappa}\Big[-\left\{\frac{1}{A^{2} B^{2}}\left(\dot F'-\frac{A'\dot F}{A}-\frac{\dot B F'}{B}\right)\right\}_{,0}\\\nonumber&&+\left\{\frac{f(R)-R F(R)}{2B^2}+\frac{\ddot F}{A^{2} B^{2}}+\frac{\dot F}{A^2 B^2}\left(\frac{\dot C}{C}+\frac{\dot D}{D}-\frac{\dot A}{A}\right)-\frac{F'}{B^4}\left(\frac{A'}{A}+\frac{C'}{C}+\frac{D'}{D}\right)\right\}_{,1}\\\nonumber&&+\frac{A'}{A B^2}\left\{\frac{\ddot F}{A^{2}}+\frac{F''}{B^{2}}-\frac{\dot F}{A^2}\left(\frac{\dot A}{A}+\frac{\dot B}{B}\right)-\frac{F'}{B^2}\left(\frac{A'}{A}+\frac{B'}{B}\right)\right\}\\\nonumber&&+\frac{2B'}{B^3}\left\{\frac{f(R)-R F(R)}{2}+\frac{\ddot F}{A^{2}}+\frac{\dot F}{A^2}\left(\frac{\dot C}{C}+\frac{\dot D}{D}-\frac{\dot A}{A}\right)-\frac{F'}{B^2}\left(\frac{A'}{A}+\frac{C'}{C}+\frac{D'}{D}\right)\right\}\\\nonumber&&+\frac{C'}{B^2 C}\left\{\frac{F''}{B^{2}}-\frac{\dot F}{A^2}\frac{\dot B}{B}-\frac{F'}{B^2}\frac{B'}{B}\right\}+\frac{D'}{B^2 D}\left\{\frac{F''}{B^{2}}-\frac{\dot F}{A^2}\frac{\dot B}{B}-\frac{F'}{B^2}\frac{B'}{B}\right\}\\\nonumber&&
-\frac{1}{A^{2} B^{2}}\left(\frac{\dot A}{A}+\frac{3\dot B}{B}+\frac{\dot C}{C}+\frac{\dot D}{D}\right)\left(\dot F'-\frac{A'\dot F}{A}-\frac{\dot B F'}{B}\right)\Big].\\\label{53}
\end{eqnarray}
\section*{Acknowledgment}
One of us G.A appreciates the financial support from HEC, Islamabad, Pakistan under NRPU project with grant number 20-4059/NRPU/R \& D/HEC/14/1217.
 
\end{document}